\documentclass[a4paper,11pt]{article}
\usepackage{pos}

\title{Axion dark matter (theory \& experiment)}

\author*[a]{Andreas Ringwald}

\affiliation[a]{Deutsches Elektronen-Synchrotron DESY,\\
  Notkestr. 85, 22607 Hamburg, Germany}

\emailAdd{andreas.ringwald@desy.de}

\abstract{We review the motivation for the axion as a solution of the strong 
CP puzzle and as a candidate for cold dark matter. Then we discuss 
benchmark axion models and present their predictions concerning 
axion couplings to the Standard Model and 
axion dark matter abundance. 
Finally, we give an overview on the discovery potential of current and planned
axion experiments, reaching from axion dark matter direct detection, 
over searches for solar axions, to direct production and detection 
of axions in the laboratory.}

\FullConference{XVIII International Conference on Topics in Astroparticle and Underground Physics (TAUP2023)\\
 28.08-01.09.2023\\
University of Vienna\\}

\begin{document}
\maketitle

\section{The Axion}

The axion appears in extensions of the Standard Model (SM) which solve two fundamental questions 
in one go: {\em i)} what is the nature of dark matter (DM), and {\em ii)} why do strong interactions obey 
$T$ and $P$ invariance so accurately? 
In fact, shortly after the proposal of Quantum Chromo-Dynamics (QCD) as the fundamental theory of 
strong interactions, it was found that its most general Lagrangian contains the so-called $\theta$-term, 
$
{\mathcal L}_{\rm QCD}\supset 
\overline\theta\,\frac{\alpha_s}{8\pi}   
  G_{\mu\nu}^b \tilde{G}^{b,\mu\nu}$,
where $\alpha_s=g_s^2/(4\pi)$ is the strong fine-structure constant, $G_{\mu\nu}^b$ ($\tilde{G}^{b,\mu\nu}$) is
the (dual) gluonic field strength, and $\overline\theta\in [-\pi,\pi]$ is an angular parameter. The $\theta$-term 
violates both $T$ and $P$ (and thus $CP$). Its contribution to the most sensitive observable of 
$T$ and $P$ violation in flavor-conserving interactions -- the electric dipole moment $d_n$ of the neutron (nEDM) -- is theoretically predicted to 
be of order 
$d_n 
\sim 10^{-16}\ \bar\theta\ e\,{\rm cm}$, while experimental measurements have up to now only established 
upper bounds on it, 
$|d_n| < 1.8\times 10^{-26}\,e\,{\rm cm}$, leading to the conclusion that $|\overline\theta| \lesssim 10^{-10}$.  

\begin{figure}[t]
\centerline{\includegraphics[width=\linewidth]{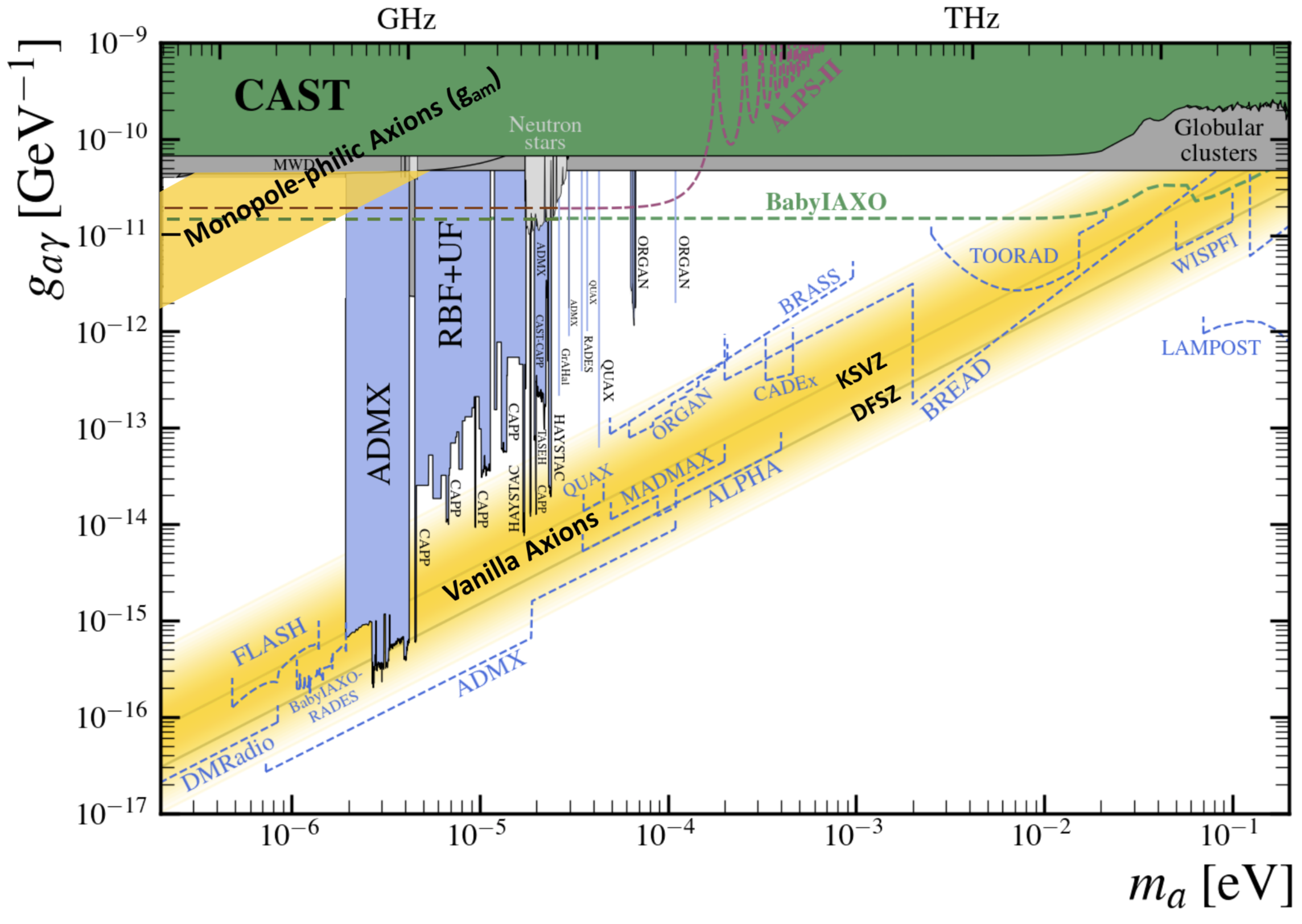}}
\caption[]{Electromagnetic coupling of the axion versus its mass (adapted from Ref.~\cite{AxionLimits}). Haloscope exclusion regions (filled blue) and projected sensitivities (dashed blue lines) assume the axion to be 100\% of the halo dark matter. In general, their sensitivities scale as  $\sqrt{\rho_{\rm DM}^{\rm halo}/\rho_a }$.}
\label{fig:electromagnetic_couplings_vs_mass}
\end{figure}

A phenomenologically viable minimal renormalizable SM extension solving the strong $CP$ problem was proposed by Kim~\cite{Kim:1979if}, Shifman, Vainshtein, Zakharov~\cite{Shifman:1979if} (KSVZ):
they added to the SM a SM-singlet complex scalar field $\sigma$, featuring a global axial Peccei-Quinn (PQ)
$U(1)$ symmetry~\cite{Peccei:1977hh}, spontaneously broken at a scale $v_{\rm PQ}$ much higher than the electroweak scale, and an exotic quark $\mathcal Q$ charged under it, 
$\mathcal{L}_{\rm KSVZ} \; \supset \; - \lambda_{\sigma} \left( \left| \sigma \right|^{2} -{v_{\rm PQ}^2}/{2} \right)^{\!\! 2} +  y \, \sigma \, \bar{\mathcal Q}_L {\mathcal Q}_R + {\rm h.c.}$. Exploiting a parametrisation of $\sigma$ in terms of a radial field $\rho$ and an angular field $a$, 
$\sigma (x) =  (1/{\sqrt{2}})\left( v_{\rm PQ} + \rho (x)\right)
{\rm e}^{i a(x)/v_{\rm PQ}}$, one sees that this model features three particles beyond the SM: two of them, $\rho$ and 
$\mathcal  Q$ are very heavy,
%
\begin{figure}[h]
\centerline{\includegraphics[width=0.44\linewidth]{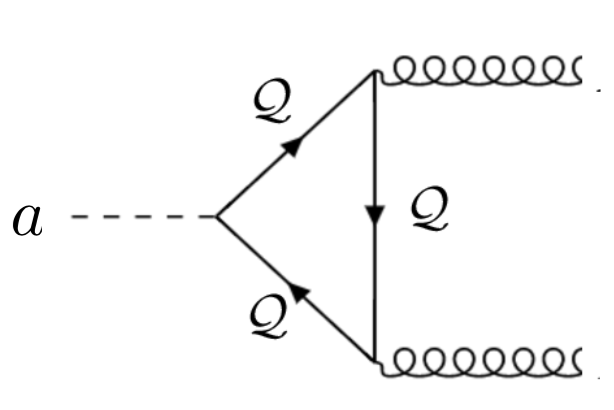}
\hspace{10ex}
                  \includegraphics[width=0.3\linewidth]{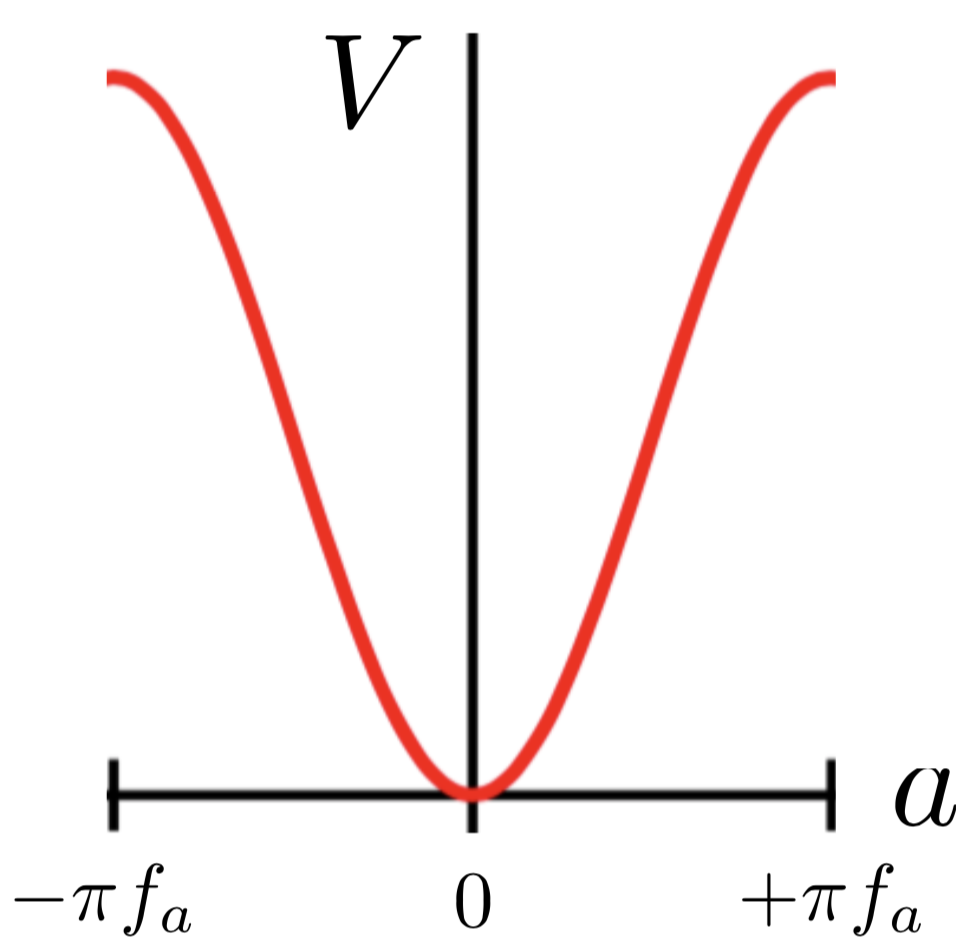}}
\caption[]{{\em Left:} Triangle loop diagram giving rise to the axion coupling to gauge fields. {\em Right:} Effective potential of the axion field.}
\label{fig:electromagnetic_coupling_axion}
\end{figure}
%
with mass of order the PQ breaking scale, $m_\rho =\sqrt{2\lambda_\sigma} v_{\rm PQ}$ and 
$m_{\mathcal Q} ={y} v_{\rm PQ}/{\sqrt{2}}$, while one, $a$, is massless: the Nambu-Goldstone boson of the $U(1)_{\rm PQ}$ breaking.
Integrating out $\rho$ and $\mathcal Q$, the dynamics of $a$ is described by
$\mathcal L_{\rm eff} 
\supset \frac{1}{2} \partial^\mu a
\,\partial_\mu a + \frac{\alpha_s}{8\pi} \frac{a}{f_a}G_{\mu\nu}^a \tilde G^{a\,\mu\nu}$. 
Here the last term arises from the triangle loop diagram in Fig.~\ref{fig:electromagnetic_coupling_axion}, which yields 
$f_a = v_{\rm PQ}/N_{\mathcal Q}$, where $N_{\mathcal Q}$ counts the number of quark flavors, which is one in the simplest case.
It has the same form as the $\theta$-term of QCD. Correspondingly, the $\overline\theta$-parameter can be eliminated by 
a shift in the $a$ field: $a (x) +  \overline\theta\,f_a \to a (x)$. Finally, taking into account the mixing of the shifted $a$ field with 
the pion field results in a non-trivial effective potential for the former which has an absolute minimum zero field value  
(see Fig.~\ref{fig:electromagnetic_coupling_axion}, right). In other words, the shifted $a$ field has a vanishing vacuum expectation value, $\langle a \rangle =0$, and thus the strong CP problem is solved~\cite{Peccei:1977hh}. 
The particle excitation~\cite{Weinberg:1977ma,Wilczek:1977pj} of the shifted $a$ field has been dubbed ``axion''. Its mass is determined by the second derivative of the axion potential 
around the minimum, 
$m_a   = {\sqrt{V^{\prime\prime}(0)}} = \frac{\sqrt{z}}{1+z}\,\frac{m_\pi \, f_\pi}{f_a} \approx 6\ {\rm \mu eV} \left( \frac{10^{12}\,{\rm GeV}}{f_a}\right)$, 
where $z=m_u/m_d\approx 1/2$ is the ratio of the masses of the up and down quark, and $m_\pi$ and $f_\pi$ are the mass and decay constant of the 
neutral pion. 
Of greatest experimental and phenomenological  interest is the axion coupling to electromagnetic interactions, 
$\mathcal L_{\rm eff} \supset \frac{1}{4} g_{a\gamma} \, a\, F_{\mu\nu} \tilde{F}^{\mu\nu}$, where 
$g_{a\gamma} 
\simeq    
 \frac{\alpha}{2\pi f_\pi } \frac{m_a}{m_\pi} 
\frac{1+z}{\sqrt{z}}
\left( {6\, q_{\mathcal Q}^{ 2}}{} - \frac{2}{3} \frac{4+z}{1+z} \right)$. 
Here,  the first term in brackets 
takes into account a possible electric charge $q_{\mathcal Q}$ of the exotic quark $\mathcal Q$ in units of $e$, 
while the second term arises from axion-pion mixing. 
The line in Fig.~\ref{fig:electromagnetic_couplings_vs_mass} labeled as `KSVZ' displays the prediction of 
$g_{a\gamma}$ for $q_{\mathcal Q}=0$. 
Varying $q_{\mathcal Q}$ gives rise to the yellow `band' of predictions~\cite{DiLuzio:2016sbl} labeled as `Vanilla Axion' in Fig.~\ref{fig:electromagnetic_couplings_vs_mass}. 
%
%
The other yellow `band' of predictions labeled as `Monopole-philic Axion' in Fig.~\ref{fig:electromagnetic_couplings_vs_mass} arises in the case that 
the KSVZ exotic quark carries a magnetic charge $g_{\mathcal Q}$, instead of an electric charge $q_{\mathcal Q}$. In this case the model would solve not only the strong CP problem, but also the charge quantisation puzzle~\cite{Dirac:1931kp,Schwinger:1966nj,Zwanziger:1970hk}. The triangle loop in
Fig.~\ref{fig:electromagnetic_coupling_axion} (left) induces then an electromagnetic  coupling~\cite{Sokolov:2021ydn,Sokolov:2021eaz,Sokolov:2022fvs,Sokolov:2023pos}
$g_{am} 
\simeq   
 \frac{\alpha_m}{2\pi f_\pi } \frac{m_a}{m_\pi} 
\frac{1+z}{\sqrt{z}} 6 g_{\mathcal Q}^2$, with
$\alpha_{m} \equiv {g_0^2}/({4\pi})$, in terms of the fundamental magnetic charge unit $g_0$.\footnote{The phenomenological viability of this non-standard electromagnetic axion coupling was recently challenged~\cite{Heidenreich:2023pbi}. We refute their arguments in an upcoming paper~\cite{Sokolov:2023toapp}.}
Charge quantisation,  $e g_0=6\pi n$, $n\in\mathbb{Z}$, results in a parametrical enhancement of the ratio of 
couplings, 
${g_{am}}/{g_{a\gamma}} = (9/4) \, \alpha^{-2}  (g_{\mathcal Q}/q_{\mathcal Q})^2 \sim 10^5$, see  Fig.~\ref{fig:electromagnetic_couplings_vs_mass}.
Such an enhancement is not possible in the axion model of  
Zhitnitsky~\cite{Zhitnitsky:1980tq} and Dine, Fischler and Srednicki~\cite{Dine:1981rt} (DFSZ model). In this model, 
the axion couplings to the gauge fields are generated by the SM fermions ($N_{\mathcal Q}=6$) which are known to 
carry no magnetic charges. The electromagnetic coupling is around the lower boundary of 
the vanilla axion band in Fig.~\ref{fig:electromagnetic_couplings_vs_mass}.

Just a few years after their original proposal it was found that axions not only solve the strong CP problem,
but are also excellent candidates for cold dark matter~\cite{Preskill:1982cy,Abbott:1982af,Dine:1982ah}.
The axion dark matter abundance prediction crucially depends on the cosmic history, in particular 
whether the PQ symmetry breaking occurs before or after the onset of the hot Big Bang phase at time $t_{\rm hot}$. 
In the former case, the so-called `pre-inflationary PQ symmetry breaking scenario', gradients in the
axion field $\theta_a (x)= a(x)/f_a$ can be neglected and its consequent evolution can be described completely by 
a damped anharmonic oscillator - the damping resulting from Hubble friction and the anharmonicity resulting from
the actual form of the periodic axion potential, which at temperatures above the QCD cross-over has the form 
$V(\theta_a,T)=(\chi (T)/f_a^2) (1-\cos\theta_a)$, where $\chi (T)$ is the topological susceptibility of QCD. 
%
\begin{figure}[t]
\centerline{\includegraphics[width=\linewidth]{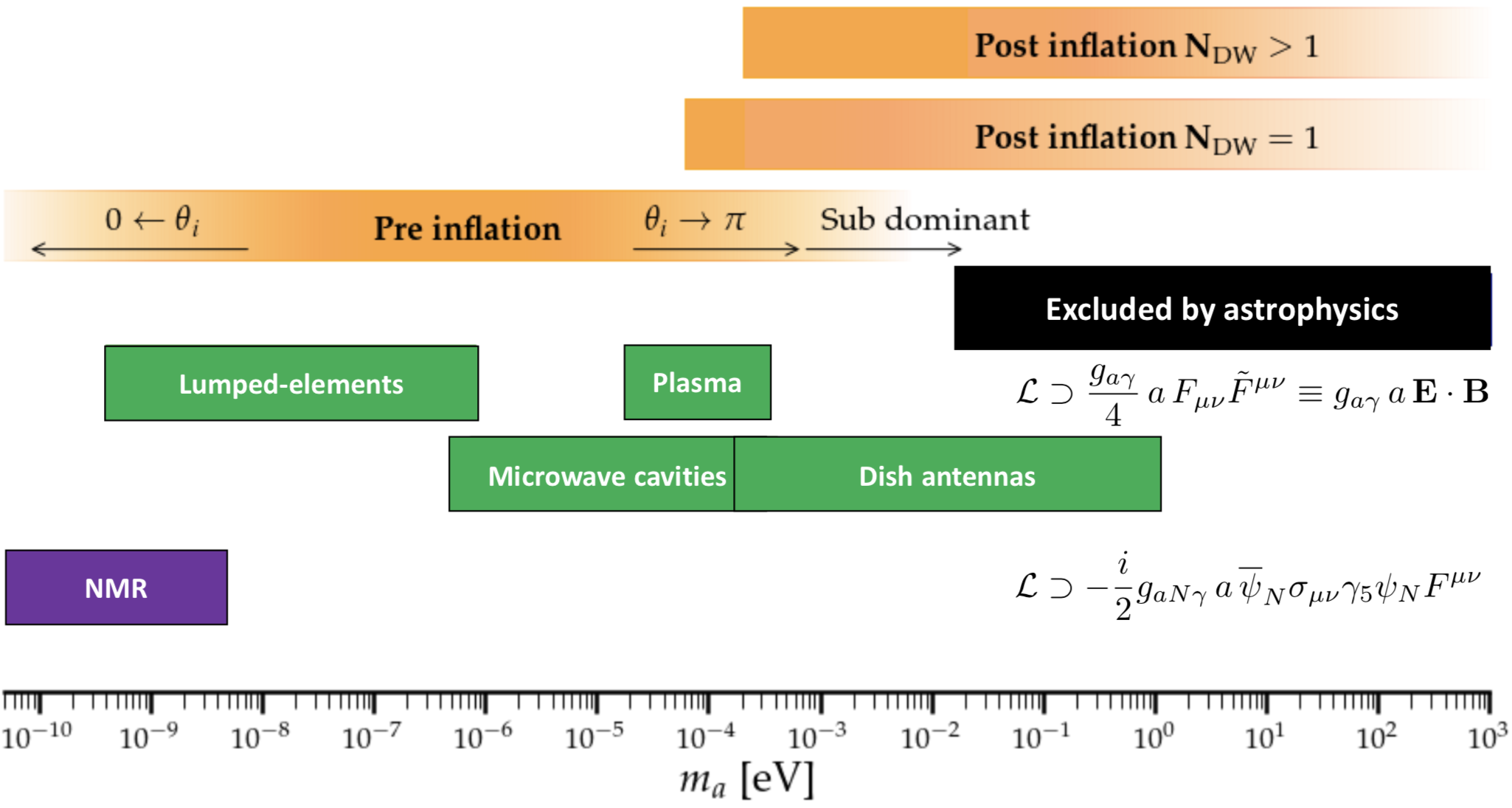}}
\caption[]{{\em Top three orange bands:} Theoretically predicted axion dark matter mass ranges.  {\em Middle black band:} Axion mass range excluded by astrophysics. {\em Green bands:} Axion dark matter ranges probed by various haloscopes. {\em Violet band:} Axion dark matter ranges probed by Nuclear Magnetic Resonance methods sensitive to NEDM oscillations or the axion dark matter wind.   Figure adapted from Ref.~\cite{Semertzidis:2021rxs}.}
\label{fig:axion_dark_matter_mass}
\end{figure}
%
The axion field is frozen at its initial value $\theta_{ai} \equiv a(t_{\rm hot})/f_a$ until the Hubble expansion rate 
drops below the axion mass $m_a(T)=\sqrt{\chi (T)}/f_a$ and oscillates thereafter around the CP conserving 
minimum. These coherent oscillations of the axion field 
ultimately represent a condensate of cold dark matter. 
The predicted axion dark matter abundance~\cite{Borsanyi:2016ksw} generated by this so-called `realignment mechanism', 
$\Omega_ah^2 
\approx 0.12\,\left({ 6~\mu{\rm eV}\over m_a}\right)^{1.165}\,
 \theta_{ai}^2$, 
depends both on the axion mass as well as on the initial value of the axion field. Therefore, 
in the pre-inflationary scenario, the requirement for axion dark matter not to 
overclose the universe does not result in a lower bound on the axion mass, see also Fig.~\ref{fig:axion_dark_matter_mass}. 
This is different in the `post-inflationary PQ symmetry breaking scenario', where $\theta_{ai}$ can take 
different values in different patches of the present universe, resulting in an
average contribution $\Delta 
 \Omega_ah^2\approx
0.12\,\left(\frac{30~\mu{\rm eV}}{ m_a}\right)^{1.165}$~\cite{Borsanyi:2016ksw} from the realignment mechanism. 
A further population of axion dark matter arises from the decay of cosmic strings and domain walls. 
Unfortunately, their predicted contribution suffers from significant uncertainties. 
Correspondingly, the plausible range of axion masses to yield 100\% of the observed dark matter in post-inflationary scenarios  is still rather large: $m_a \approx 26\ \mu{\rm eV}  - 0.5\ {\rm meV}$, for axion models with short-lived domain walls~\cite{Klaer:2017ond,Gorghetto:2020qws,Buschmann:2021sdq}, such as the
KSVZ model ($N_{\rm DW}\equiv N_{\mathcal Q}=1$). For models with long-lived domain walls ($N_{\rm DW}>1$), such as a DFSZ model where the PQ symmetry arises from an accidential discrete symmetry~\cite{Ringwald:2015dsf}, 
the mass is predicted to be significantly higher~\cite{Ringwald:2015dsf,Beyer:2022ywc}: 
$
m_a \gtrsim {\rm meV}$.
Large density variations in the initial state of the axion field in the post-inflationary scenario lead to the formation of compact dark matter objects known as “miniclusters”. These objects  lead to an increased theoretical uncertainty in the local axion density for dark matter detection in the laboratory.

%
\begin{figure}[t]
\centerline{\includegraphics[width=\linewidth]{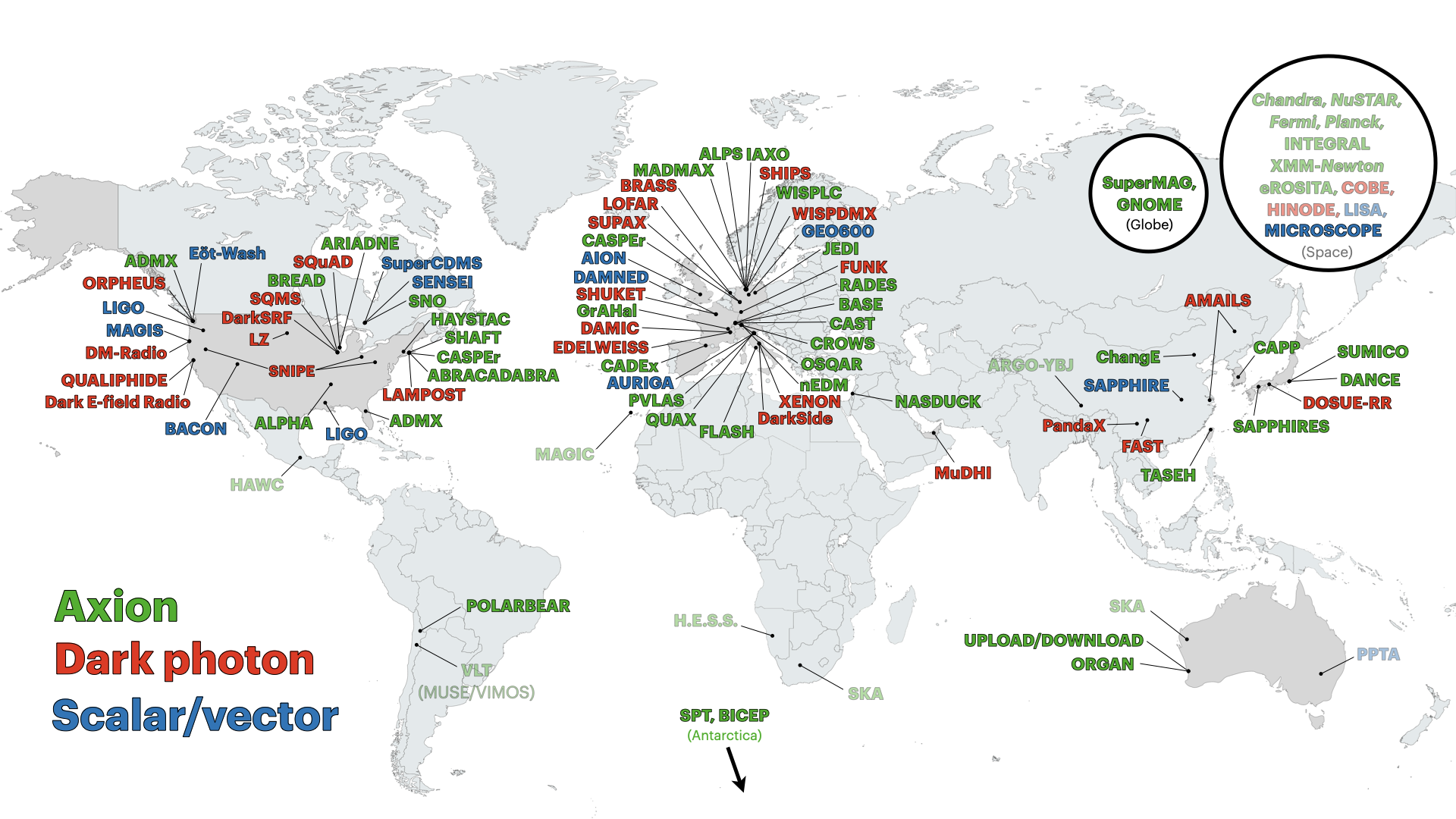}}
\caption[]{World map displaying current experiments searching for wavy dark matter~\cite{AxionLimits}.}
\label{fig:wavy_dark_matter_exp_world_wide}
\end{figure}
%

\section{Axion Dark Matter Experiments}

The current predictions of the axion mass possibly explaining the observed amount of dark matter span many orders of 
magnitude (see Fig.~\ref{fig:axion_dark_matter_mass}). Correspondingly, various different experimental techniques, exploiting 
even different couplings,  are required to explore this vast parameter space.  An enormous number of axion experiments is
presently being pursued world-wide, see Fig.~\ref{fig:wavy_dark_matter_exp_world_wide}.  Clearly, I can only concentrate on a subset of them in the remaining part of this short review.

Axion dark matter experiments rely on the assumption that the dark matter halo of the Milky Way is comprised of axions. 
Their velocity dispersion is given by the galactic virial velocity, implying a macroscopic de Broglie wave length, 
$\lambda_{\rm dB} = 2\pi/(m_a v_a)\simeq {\rm km}\, (\mu{\rm eV}/m_a)(10^{-3}\,c/v_a)$. Correspondingly, 
halo dark-matter axions behave as an approximately spatially homogeneous and monochromatic classical oscillating field,
$a(t) \simeq \sqrt{2\rho_a} \cos (m_a t)/m_a$: they belong to the category of `wavy dark matter'. 
When presenting (projected) limits on halo dark-matter axion couplings it is assumed that
$\rho_a = \rho_{\rm DM}^{\rm halo} \approx 0.45\,{\rm GeV}\,{\rm cm}^{-3}$.

A big number of axion dark matter experiments relies on the electromagnetic coupling, 
${\mathcal L} \supset \frac{g_{a\gamma}}{4}\, a\, F_{\mu\nu} \tilde F^{\mu\nu} \equiv g_{a\gamma}\,a\,{\mathbf E}\cdot {\mathbf B}$, 
and are therefore called `haloscopes'. Haloscopes exploit different experimental techniques which are suited 
in different mass ranges, see the green areas in Fig.~\ref{fig:axion_dark_matter_mass}. 

The original classic haloscope concept was proposed by Sikivie~\cite{Sikivie:1983ip}: a microwave cavity is placed in a magnetic field, in which the dark matter axion converts in a photon. 
If the  axion mass matches the resonance frequency of the cavity, $m_a = 2\pi \nu_{\rm res} \sim 4\,\mu{\rm eV} \left( \frac{\nu_{\rm res}}{\rm GHz}\right)$, the power output is enhanced by its quality factor, $P_{\rm out} \sim g_{a\gamma}^2\, \rho_{\rm a}\, B_0^2\, V\, Q$. 
Since the axion mass is not known one needs to scan by tuning the resonance frequency. There are many microwave cavity 
haloscopes running at the moment, two of which, namely ADMX~\cite{ADMX:2019uok} and CAPP~\cite{Yi:2022fmn}, have already reached nominal DFSZ sensitivity 
in some mass range, see Fig.~\ref{fig:electromagnetic_couplings_vs_mass}. 
Within the next decade, microwave cavity axion dark matter searches, such as e.g. ADMX~\cite{Stern:2016bbw}, BabyIAXO~\cite{Ahyoune:2023gfw}, and FLASH~\cite{Alesini:2023qed}, will dig deeply into the vanilla axion band in the range $0.5\,{\rm \mu eV} \lesssim m_a \lesssim 100\, {\rm \mu eV}$, cf.  Fig.~\ref{fig:electromagnetic_couplings_vs_mass}. 
Unfortunately, these searches are insensitive to the non-standard coupling $g_{am}$ of the monopole-philic axion~\cite{Tobar:2022rko}.

Higher masses can be probed by a broadband search method based on the dish antenna haloscope concept~\cite{Horns:2012jf}. 
It relies on the fact that oscillating axion dark matter, in a background magnetic field $\mathbf B$, carries an oscillating electric field component parallel to the latter, ${\mathbf E}_a(t) = - g_{a\gamma}{\mathbf B} a(t)$. Correspondingly, a metallic mirror placed in a magnetic field pointing parallel to the mirror surface will emit a
nearly monochromatic 
EM wave perpendicular to the mirror surface with a frequency $\nu = m_a/(2\pi)$ and a cycle-averaged 
power per unit area, ${{\mathcal P}_\gamma}/{\mathcal A}=|\mathbf{E}_a|^2/2$. 
The BRASS collaboration at the University of Hamburg is currently setting up a pilot dish antenna haloscope exploiting a permanently magnetized conversion panel~\cite{Bajjali:2023uis}, while the 
BREAD collaboration at Fermilab plans to exploit a cylindric parabolic conversion panel which allows to use a much stronger 
solenoidal magnetic field~\cite{BREAD:2021tpx}. In their final stage, these experiments have the potential to scratch the vanilla axion band region 
in the range $50\,{\rm \mu eV} \lesssim m_a \lesssim {\rm meV}$ (BRASS) or to dive deeply into it for 
$20\,{\rm meV} \lesssim m_a \lesssim 0.1\,{\rm eV}$ (BREAD), see Fig.~\ref{fig:electromagnetic_couplings_vs_mass}. 

An alternative technique to search for axion dark matter in the mass range above $\sim 50\,{\rm \mu eV}$ is based 
on the concept of a dielectric haloscope~\cite{Caldwell:2016dcw}, which is essentially a boosted dish antenna haloscope:
it consists of a mirror and a series of parallel, partially transparent dielectric disks in front of it, all placed within a magnetic field parallel to the surfaces, and a receiver in the field-free region. 
Each disk acts as a flat dish antenna. The waves emitted by each disk are reflected by and transmitted through the other disks before exiting. For suitable disk separations, these waves add coherently to enhance the emitted power.    
This allows scans over a band of $m_a$ without needing to use disks with different thicknesses for each measurement. 
Based on this concept, the MADMAX collaboration aims at installing at DESY in Hamburg an 
adjustable multiple-disk system (booster) of $\mathcal A \sim  1$\,m$^2$ inside a $\sim 9$\,T dipole magnet~\cite{MADMAX:2019pub}. 
Data taking in Hamburg may start in 2030, but already earlier a prototype magnet could allow for axion dark matter 
searches in mass and frequency ranges which are so far largely unexplored.
With an expected power boost factor of $\beta^2(\nu )\sim 10^4$ and equipped with a quantum-limited receiver, 
the full MADMAX experiment is projected 
to scan the $(40-400)\,\mu$eV mass range with DFSZ sensitivity~\cite{Beurthey:2020yuq}, as shown in Fig.~\ref{fig:electromagnetic_couplings_vs_mass}. 

A third alternative technique suitable in this mass range is based on the plasma haloscope concept~\cite{Lawson:2019brd}: 
it relies on the fact that, in a magnetized plasma, the oscillating axion dark matter field induces plasmon excitations,
${\bf E}  
=-g_{a\gamma}{\bf B}_{\rm e}a \left(1-\frac{\omega_p^2}{\omega_a^2-i\omega_a\Gamma}\right)^{-1}$. 
The latter are resonantly enhanced, when the plasma frequency matches the axion mass, $\omega_p = \omega_a \approx m_a$,
limited by losses ($\Gamma$). 
A plasma with tunable plasma frequency in the GHz range can be realised by a wire array with variable interwire spacing (“wire metamaterial”).
The ALPHA collaboration aims to build at Oak Ridge National Laboratory a  tunable cryogenic plasma haloscope along these lines. 
Presently a pathfinder experiment is developed which should be taking data in 2026. The 
full ALPHA experiment is designed to dig deep into the vanilla axion band in the $(40-400)\,\mu$eV mass range, see  Fig.~\ref{fig:electromagnetic_couplings_vs_mass}. 

The concept of a lumped element axion haloscope, sensitive in the sub-micro-eV mass range, was proposed about ten years ago~\cite{Sikivie:2013laa}: it is based on the fact that, 
in the presence of a  
magnetic field $\mathbf B$, the axion dark matter field induces an oscillating effective displacement 
current, ${\mathbf j}_a = - g_{a\gamma} {\mathbf B} \dot a$, which in turn generates an  
oscillating magnetic field 
${\mathbf B}_a$,  such that ${\mathbf\nabla}\times {\mathbf B}_a = {\mathbf j}_a$. 
The induced field can be turned into 
an alternating current in a pickup loop, resonantly amplified in a tunable LC circuit, and finally detected via a SQUID. 
There have been already pilot experiments along these lines (ABRACADABRA~\cite{Ouellet:2018beu,Salemi:2021gck}, 
ADMX SLIC~\cite{Crisosto:2019fcj}, SHAFT~\cite{Gramolin:2020ict}), and the next 
generation is currently starting (WISPLC~\cite{Zhang:2021bpa}, DMRadio~\cite{DMRadio:2022pkf}). The ultimate goal of the DMRadio collaboration is to probe 
axion dark matter in the range from neV to micro-eV down to DFSZ sensitivity, see Fig.~\ref{fig:electromagnetic_couplings_vs_mass}.
A variant lumped-element haloscope exploiting a high-voltage capacitor instead of a magnetic field may also probe the $g_{am}$ coupling of the monopole-philic axion~\cite{Li:2022oel,Tobar:2023rga}. 

Let us finally mention an axion dark matter experiment not relying on the electromagnetic coupling, but 
rather on the least model dependent coupling of the axion: the coupling to the nucleon EDM operator,  
$\mathcal{L}_{aN\gamma} = -\frac{i}{2} g_{aN\gamma}\, a\, \overline{\psi}_N \sigma_{\mu\nu} \gamma_5 \psi_N F^{\mu\nu}$, 
where $g_{an\gamma}=  - g_{ap\gamma}
\approx 
6\times 10^{-19} \left( \frac{m_a}{\rm neV}\right) \frac{1}{\rm GeV^2}$. 
The oscillating halo dark matter axion field induces oscillating nuclear electric
dipole moments (EDMs),
$
d_N (t) = g_{aN\gamma}\,\sqrt{2\rho_{a}} \cos (m_a\,t)/m_a$~\cite{Graham:2013gfa}.  
These cause the precession of nuclear spins in a nucleon spin polarized
sample in the presence of an electric feld. The
resulting transverse magnetization can be searched for by exploiting nuclear magnetic resonance (NMR)
techniques, which are most sensitive
in the range of low oscillation frequencies corresponding to sub-neV axion masses, see violet band in Fig.~\ref{fig:electromagnetic_couplings_vs_mass}. 
CASPEr-electric~\cite{Budker:2013hfa} in Boston is developing such an NMR axion search experiment~\cite{Aybas:2021nvn}.
Its final goal is to probe axion dark matter with neV masses, corresponding to a PQ scale of order 
$10^{16}$\,GeV, 
as predicted in Grand Unified models (see e.g.~\cite{Ernst:2018bib,DiLuzio:2018gqe,FileviezPerez:2019fku}). 
  
\section{Experimental Searches for Home-Made and Solar Axions}

To complete the landscape of axion experiments, let me mention here also two experimental 
techniques which are not relying on axions being the main constituents of the halo dark matter. 

A powerful tool to search for axions in a pure laboratory and thus astrophysical-model independent setup is the 
``Light Shining through a Wall" (LSW)  technique~\cite{Anselm:1985obz,VanBibber:1987rq}.  
It is based on the fact that photons sent along a transverse magnetic field $B$ of length $L_B$ may 
convert partially into light axions, and vice versa, with a probability\footnote{Here, $g_{a\gamma}$ can be replaced by $g_{am}$ for the monopole-philic axion~\cite{Sokolov:2022fvs}.}  
$P(\gamma \rightarrow a ) \simeq \frac{1}{4} \left( g_{a\gamma} B L_B\right)^2\simeq P(a \rightarrow \gamma )$. 
Correspondingly, putting a light-tight wall in the middle of the transverse magnetic
field region, the existence of axions is signaled by photons emerging behind the wall from 
axion-photon conversion. 
The LSW experiment ALPS II~\cite{Bahre:2013ywa}  at DESY in Hamburg 
exploits 
{\em i)} two strings constituted by recycled superconducting HERA dipole magnets -- 12 before and 12 after the wall -- 
in one of the straight sections of the HERA tunnel~\cite{Ringwald:2003nsa} instead of only one such magnet as in its preceding ALPS~\cite{ALPS:2009des,Ehret:2010mh} experiment and {\em ii)} an optical cavity also on the after-wall side 
to enhance resonantly their number~\cite{Hoogeveen:1990vq} instead of using only one optical cavity before the wall as in ALPS. 
ALPS II data taking has started in May 2023 and is expected to reach its full sensitivity in 2025. 
It will be about  
$10^3$-times more sensitive than the last generation of LSW experiments (ALPS, OSQAR~\cite{OSQAR:2015qdv}) and 
probe previously uncharted  
parameter space, in particular the monopole-philic axion, cf.   Fig.~\ref{fig:electromagnetic_couplings_vs_mass}.

The production of axions through the interaction of photons with the Coulomb-field of nuclei in the solar plasma may lead to a sizeable flux of solar axions, which can be searched for at Earth with an  
axion helioscope~\cite{Sikivie:1983ip}: a long dipole magnet pointed towards the 
sun in which solar axions are partially converted into photons which can be focused then on an  X-ray detector.   
CAST at CERN has established an upper limit~\cite{CAST:2017uph} on $g_{a\gamma}$, resp. $g_{am}$, which, for 
$m_a\lesssim 10$\,meV, nearly coincides with the limit from stars in globular clusters in Fig.~\ref{fig:electromagnetic_couplings_vs_mass}. 
The next generation helioscope will be BabyIAXO~\cite{IAXO:2020wwp} at DESY in Hamburg. 
It will exploit a 10\, m long superconducting magnet ($\sim 2$\,T) with two bores, each with a diameter of 70\,cm. 
The two detection lines will feature both X-ray optics and an ultra-low background X-ray detector.
BabyIAXO is designed to exceed the sensitivity of CAST on $g_{a\gamma}$, resp.  $g_{am}$, by a factor of around four in 
the same axion mass range (see Fig.~\ref{fig:electromagnetic_couplings_vs_mass}). 
It will probe also the $\gtrsim 10$\,meV vanilla axion. 
Data taking may start in 2028.

\section*{Acknowledgments}
Special thanks to A.~Lindner, G.~Raffelt, C.~Schwemmbauer, and A.~Sokolov for valuable comments on the draft.
This work has been partially funded by the Deutsche Forschungsgemeinschaft (DFG, German Research Foundation) 
under Germany’s Excellence Strategy - EXC 2121 Quantum Universe - 390833306  and under 
- 491245950.
This article/publication is based upon work from COST Action COSMIC WISPers CA21106, supported by COST (European Cooperation in Science and Technology).

\end{document}